# Evaluation of the number of undiagnosed infected in an outbreak using source of infection measurements

Akiva B. Melka[1] and Yoram Louzoun[1,2,*]

[1]Department of Mathematics, Bar-Ilan University, Ramat Gan 52900, Israel

[2]Gonda Brain Research Center, Bar-Ilan University, Ramat Gan 52900, Israel

[*]Corresponding author: louzouy@math.biu.ac.il

## ABSTRACT

In times of outbreaks, an essential requirement for better monitoring is the evaluation of the number of undiagnosed infected individuals. An accurate estimate of this fraction is crucial for the assessment of the situation and the establishment of protective measures. In most current studies using epidemics models, the total number of infected is either approximated by the number of diagnosed individuals or is dependent on the model parameters and assumptions, which are often debated. We here study the relationship between the fraction of diagnosed infected out of all infected, and the fraction of infected with known contaminator out of all diagnosed infected. We show that those two are approximately the same in exponential models and across most models currently used in the study of epidemics, independently of the model parameters. As an application, we compute an estimate of the effective number of infected by the SARS-CoV-2 virus in various countries.

## Introduction

In the absence of a vaccine or efficient treatment, the control of social contacts through large-scale social distancing measures appears to be the most effective means of mitigation in a pandemic[1–5]. Determining the extent of those measures and their stringency requires an accurate evaluation of the total number of infected individuals along with the fraction of those individuals that have not yet been identified[6–8]. Many parameters can influence this evaluation. For instance, when a disease or a virus has a short incubation period and a relatively small spreading rate compared to its detection rate, the fraction of undiagnosed infected is relatively small and the outbreak can be stopped or, at the least, contained, by isolating the infected individuals from the population[9]. In opposite cases, such as in the HIV, SARS, EBOV, or SARS-CoV-2 outbreaks, the fraction of undiagnosed infected can be substantial, and spreading can occur through them[10–12]. Modeling has emerged as an important tool in determining the effectiveness of those measures. It enables to gauge the potential for widespread contagion, cope with associated uncertainty, and inform its mitigation[13–15].

To estimate the total number of infected from observed infected, one needs to determine the Confirmed Cases Fraction (*CCF*), defined here as the fraction of confirmed (diagnosed) infected out of all infected (both diagnosed and undiagnosed). The reported number of carriers is heavily influenced by sampling biases. this number is usually incomplete due to the lack of testing capacities, and varying testing protocols[16,17]. We here propose that *CCF* can be estimated through the Known Source Fraction (*KSF*), defined as the fraction of diagnosed individuals with known contaminators. Epidemiological investigations, even on a limited sample of the confirmed infected individuals, can provide the value of *KSF* and therefore an estimation of *CCF*.  Such a sample would have to be sufficiently diversified to represent the population, and especially the variability in infection probability (e.g., the difference between super-spreaders ad regular spreaders). We show that even in a population with diverse infection rates, *KSF* provides a rather accurate estimation of *CCF,* for an unbiased sample above a minimal size.

In contrast with *KSF*, *CCF* can only be directly measured through wide scales surveys. Moreover, the total fraction of infected is usually low, requiring very large surveys to obtain accurate estimates of *CCF*.

## Results

Two main types of predictive models were proposed for epidemics: macroscopic models, using aggregated data at the population scale, and microscopic models, incorporating distributed information at the individual level[18,19]. Macroscopic models use stochastic processes or ODEs to predict the evolution of the outbreak on a global scale. The simplest and most common model is the SIR model[20,21], where the population is divided into three categories: Susceptible (*S*), Infected (*I*), and Removed (*R*) (Fig. 1a). *N* is the total population.

In this model, propagation of the virus depends on the infection rate $\beta$ or the number of contacts between susceptible and infected individuals, and the detection rate $\gamma$ that characterizes the time that infected individuals remain contagious. The Removed category can include individuals that survived the virus and are now immune or deceased patients. If stringent confinement is applied, this category can also simply be all diagnosed individuals since they are now removed from the system and can no longer contaminate other individuals. To model *KSF*, we add a category to a stochastic realization of SIR and other models: Controlled (*C*) that represents the individuals among the Removed for whom the contaminator is known. In practice, each time a Susceptible gets infected, an Infected is chosen to be the contaminator and its identity is recorded. When an individual gets diagnosed, we check the identity of its contaminator and if this contaminator has already been diagnosed, we consider that the newly diagnosed individual is added to the Controlled category (see Fig. 1c for a description). We ignored false positives (diagnosed that are not infected) in the current analysis, as their number is consistently small in most epidemics[22]. We further discuss false negatives.

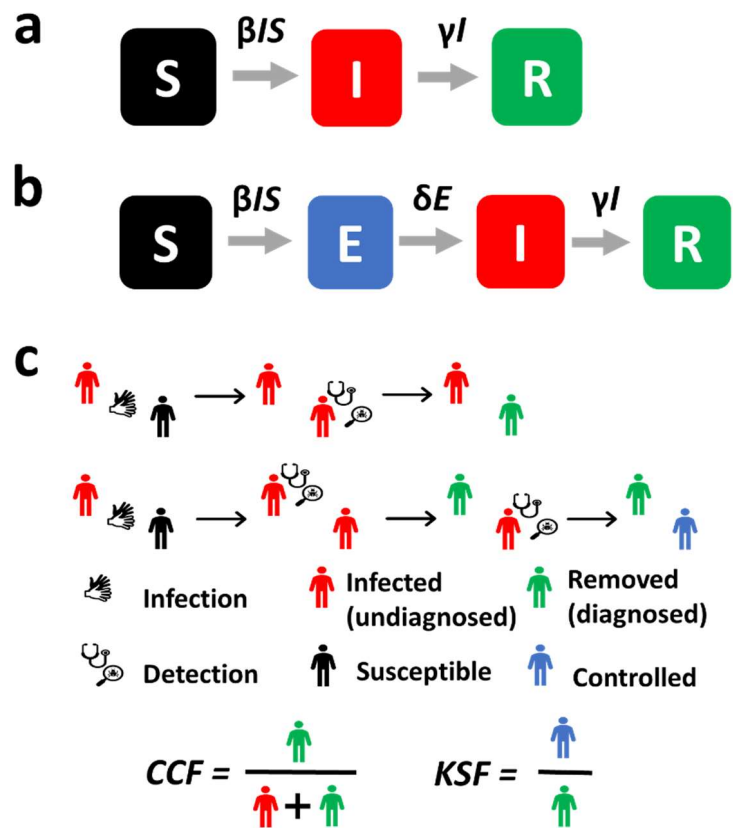

**Figure 1.** Models Description. (**a**) dynamics of the SIR model. A Susceptible individual can get infected with a rate proportional to $\beta IS$. An infected can get Removed from the system with a rate proportional to $\gamma I$. (**b**) dynamics of the SEIR model. An Exposed category is added. Exposed are not infecting but can become infecting with a probability of $\delta$ per exposed. (**c**) dynamics of the discrete-time simulations: First line, each Infected (red) can infect each Susceptible (dark). If an infected is detected, it becomes quarantined and thus removed (green). Second line, If the contaminator of a diagnosed individual was already detected (i.e., it is green by the time the new Infected is diagnosed), the newly diagnosed is considered Controlled (blue) implying that its source of contamination is known. We define two ratios. *CCF* is the fraction of diagnosed individuals over the total number of infected (diagnosed and undiagnosed). *KSF* is the fraction of diagnosed individuals with a known source of contamination.

The first order average dynamics of the SIRC model can be approximated by:

$$(1) \quad \frac{dS}{dt} = -\beta S \frac{I}{N}, \quad \frac{dI}{dt} = \beta S \frac{I}{N} - \gamma I, \quad \frac{dR}{dt} = \gamma I, \quad \frac{dC}{dt} = \frac{R}{I+R} \gamma I, \quad \text{and} \quad S + I + R = N.$$

Assuming that initially *S = N*, *R = C* = 0, and *I* is very small, and solving for small variations (see the Methods section for derivations) one gets that *CCF = KSF*.

As mentioned above, the evaluation of *CCF* is most relevant when the incubation period is significant. For example, in the current COVID-19 pandemic, a lag of a few days has to be considered when observing infection patterns[23]. Therefore, we also extended the SEIR model to a SEIRC model. In this model, an Exposed (*E*) category is added that represents the infected individuals that carry the virus but still do not contaminate others (Fig. 1b). $\delta$ is the rate at which an Exposed becomes infectious and can now contaminate other individuals. The dynamics equations are

$$(2) \quad \frac{dS}{dt} = -\beta S \frac{I}{N}, \quad \frac{dE}{dt} = \beta S \frac{I}{N} - \delta E, \quad \frac{dI}{dt} = \delta E - \gamma I,$$

$$(3) \quad \frac{dR}{dt} = \gamma I, \quad \frac{dC}{dt} = \frac{R}{I+R} \gamma I, \quad \text{and} \quad S + E + I + R = N.$$

With the same assumptions, we also get that, up to a short and small transient, *CCF = KSF*. This can be easily derived from the equations since *KSF = C / R* and *CCF = R / (I + R)*. This equality holds for any model producing an exponential growth of *I, R,* and *C*. *CCF* and *KSF* are ratios of population fractions that grow with the same exponent. Thus, the exponent functions cancel out in the denominator and numerator, leading to constant values. Moreover, in exponential growth models[14] (as is observed in the real-world at the early stage of most epidemics), we can add that, not only *KSF* and *CCF* are constant, but also equal (see the Methods section).

While there is a large number of existing epidemiological models for any pandemic, and specifically for the current COVID-19 pandemic, most current works are based on different versions of either the SIR or SEIR models[23]. More sophisticated versions of SEIR also incorporate migration to assess the efficiency of intercity restrictions[24] or other categories such as asymptomatic individuals[25]. Finally, models were refined with a time-dependent detection rate, age-dependent infection matrices[26], or even quarantine.[27,28]. However, the vast majority of these models produce an initial exponential growth, and as such, we expect the equivalence between *KSF* and *CCF* to hold.

To validate this equivalence, we tested multiple models. In most realistic cases, the spread dynamics parameters or even the appropriate model are unknown. To show that the relationship between *KSF* and *CCF* is not model or parameter specific, we tested this relationship in multiple versions of SIR and SEIR models with different parameter configurations (see Fig. 2c for the results and the Methods section for a description of the different sets of parameters and simulations). We implemented SIR and SEIR models with homogenous and heterogeneous infection rates to reflect the fact that not all individuals have the same infection probability (as a function of age/gender/genetics or other factors). We used a power-law distribution for the infection rate, as is most commonly observed across populations. The variability in the infection probability can be modeled either as a variability in the probability of each individual to infect others or in the probability of each individual to be infected. We modeled both types of variability. Although the presence of those super spreaders/high-risk infected might locally disturb the distribution or the growth, on a global scale, the relation between *CCF* and *KSF* still holds. We also tested a time-dependent detection rate (Fig. 2e, f). If the increase in detection rate is relatively slow (compared with the exponent of the infected class size), we still observe an exponential growth, and our results hold. One can see that, while both fractions vary in different models and parameters, however, an approximately linear relationship is consistent among all models. Besides, those two fractions rapidly achieve equilibrium (Fig. 2a, b). Moreover, in different realizations of the same model, most of the trajectory density is centered on a limited range of *KSF* and *CCF* values. Different initial conditions and stochastic realizations lead to similar solutions (Fig. 2b). As such, one can use *KSF* to estimate *CCF* without further knowledge of the model or its parameters. As observed in Fig. 2d, the ratios remain close, even after the initial phase. This happens since the two ratios were equal in the initial phase and then both tend to 1. Although the numerical results suggest that even after the initial phase, *KSF* is still close to *CCF*, this may depend on the details of the model.

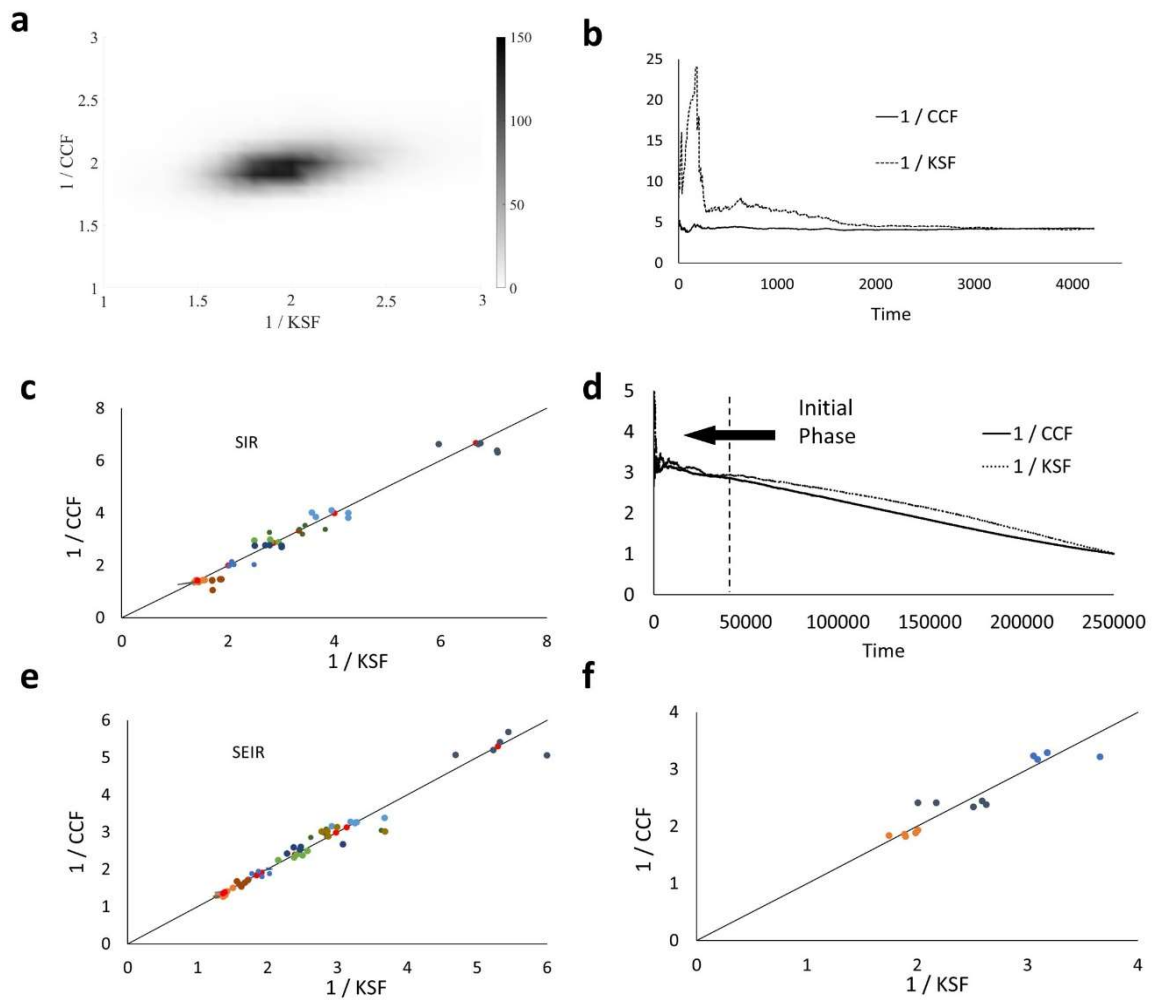

**Figure 2.** Results from simulations. (**a**) density plot of the two ratios over 1,000 simulations. (**b**) time evolution of the two ratios $\frac{1}{CCF} = 1 + \frac{I}{R}$ and $\frac{1}{KSF} = \frac{R}{C}$. We observe that not only do the ratios achieve equilibrium, but they are never very far from it. (**c**) Simulations of the SIR model with different sets of parameters and variable infection rate. Red dots correspond to the analytical values. (**d**) Beyond the initial phase, we observe that *CCF* and *KSF*, although still close, part and then both tend to 1. (**e**) Simulations of the SEIR model with different sets of parameters and variable infection rate. Red dots correspond to the analytical values. (**f**) Simulations of the

SIR model. One set is with a time-dependent detection rate, one with the presence of super spreaders, and one with both a variable infection rate and super spreaders. In all models used (SIR or SEIR, with homogeneous or heterogeneous infection rate or time-dependent detection rate), we observe a linear relationship between *1/CCF* and *1/KSF* in the growth phase.

At the practical level, since *R* and *C* can be obtained from measures of diagnosed infected and epidemiological investigations, *KSF* can be estimated in most cases. Then *CCF* and thus *I* can be determined from the equivalence. To check the applicability of our methodology, we analyzed the number of confirmed cases (*CCF*) for the COVID-19 pandemic and the fraction of confirmed cases with known sources (*KSF*) in Israel, Mexico, the Philippines, and Hong Kong. We then estimated the total number of infected (Fig. 3). The choice of those countries and the period for which we computed our estimate were determined by the published data, but this could easily be performed for any other country.

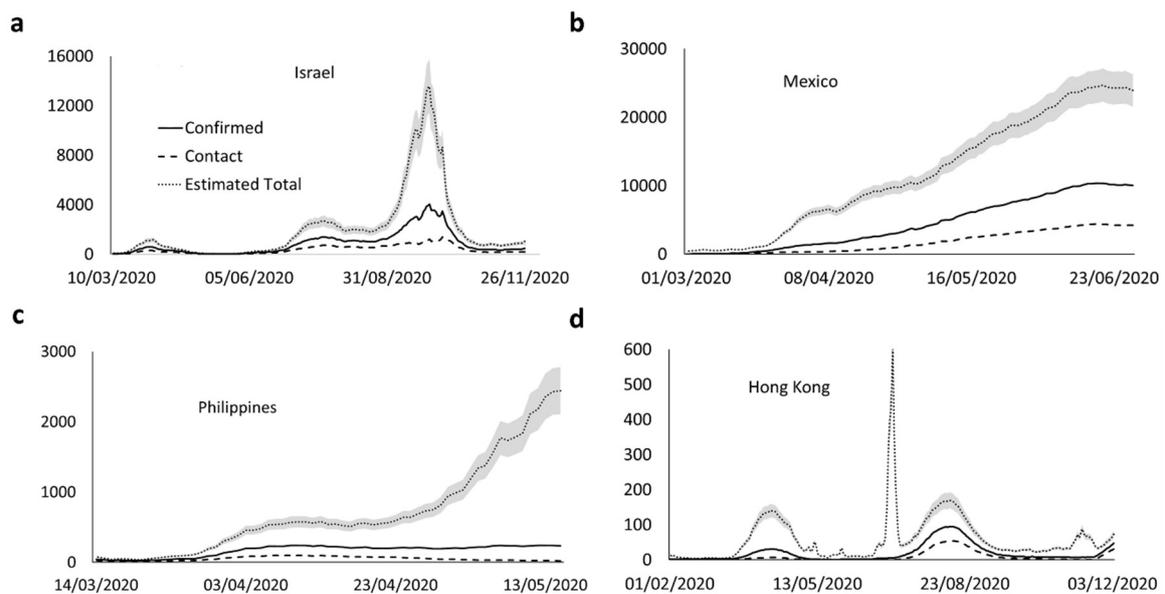

**Figure 3.** Estimation from real-world data. Daily number of infections for different countries. (**a**) Israel, (**b**) Mexico, (**c**) the Philippines, and (**d**) Hong Kong. The full line represents the number of infected diagnosed daily, the dashed line represents the number of diagnosed with a known source, the dotted line represents the estimation of the total number of infected, and the gray shading represents the 95% confidence interval of the error induced by the estimate. All curves were computed with a moving seven-day average. We observe in the Hong Kong curve a limitation of the method: a drop in the fraction of cases with known origin reported can lead to a sharp rise in the number of estimated infected.

**Discussion**

Multiple models have been proposed to evaluate *CCF* using, for instance, the number of deceased patients[29,30], but in all those studies, the results depend on the models used or on estimates of country-specific parameters, such as the age dependence or the Infection fatality rate. We have presented a method to estimate the fraction of undiagnosed infected from the fraction of infected with a known contaminator (out of all infected). While the first value is hard to measure in realistic situations, the second is often known. Our rationale is that each infected has a given probability to be detected, approximated by the fraction of the confirmed cases (*CCF*). Similarly, for each infected, there is a given probability to determine who caused the infection, approximated by the fraction of infected with an identified source (*KSF*). We proved that not only those two probabilities are related but they are equal during the exponential growth phase of an epidemic.

The *KSF* estimate suffers from multiple caveats with opposite effects. First, removed individuals are considered controlled only if their contaminators were already diagnosed when in fact it could be diagnosed even after. Therefore, even already removed individuals could be counted eventually as controlled. A second and more complex problem is that reported infected may be biased toward people who have been in contact with other reported infected. As such, the number of controlled individuals would be overestimated. As mentioned above, the presence of super spreaders would also bias the computation of *KSF* in very small samples and alter the estimation of *CCF*[31,32]. A direct solution to these limitations would be to perform detailed epidemiological investigations on patients with clinical complications. Such patients typically do not suffer from sampling bias and detailed enough investigations will limit the number of missed controls. Such investigations can be performed on a limited sample[33,34]. Another limitation of our estimate is that epidemiological investigations are not perfect, as such, some controlled individuals might be missed. Similarly, some diagnosed may be assumed to be infected from a known source, when in fact they were infected by other sources. These limitations can be solved when detailed genetic information is available on the virus or disease. Note again that only a small fraction of the diagnosed individuals needs to be investigated in detail to obtain *KSF* as long as this fraction is unbiased.

Other versions of the SIR models include a transition to a death state or an Asymptomatic category[35]. Since our Removed category includes all individuals that were diagnosed and, hence, can no longer contaminate, it already accounts for the dead and the effect of quarantine. Besides, our Infected category includes all undiagnosed individuals that can still contaminate others therefore, it accounts for all carriers including the asymptomatic individuals. These asymptomatic are indeed included in the category of individuals for whom we have no information (they have not been tested or identified yet). In the COVID-19 pandemic, migration has a minor effect on contamination[24], so we did not include it. However, in the presence of significant migration, the model presented here will not be valid. For the sake of simplicity, we presented here a non-spatial model where all infected individuals can contaminate others disregarding proximity but, since the similarity between *CCF* and *KSF* is an inherent property of epidemiological models, we do not expect network and spatial features to change our conclusions.

To summarize, as is the case for every model, multiple caveats can affect the validity of the model, most of those can be avoided in detailed and unbiased investigations on small numbers of diagnosed (even a few tens). Thus, while the efficacy of contract tracing has been extensively studied and debated[36,37], we do not propose to use the observed relationship as is on biased published epidemiological data. The here reported relationship between *KSF* and *CCF* can be a critical tool to estimate the spread of diseases. A minimum tracing effort (or record of the individuals that were already in quarantine before positive testing) is of course necessary. In the absence of tracing, *KSF* would not be available leading to a miscalculation of *CCF*. This explains the glitch in the Hong Kong graph in Fig 3. It must be noted that this happens during a relaxation period with almost no newly diagnosed individuals and therefore no tracing performed. Even if tracing efforts can reveal insufficient but still minimal, we claim that gaps in tracing reflect gaps in testing, and just as the fraction of traced diagnosed individuals would be low, so would the fraction of diagnosed versus the actual number of infected people. It might also be added (as is observed on the data from different countries) that tracing was reinforced during the second wave of the COVID-19 pandemic demonstrating how health organizations redeem it relevant.

## Methods

**SIR Model.** The equations for the SIR model can be simplified by using

$$(4) \quad s = \frac{S}{N}, \quad i = \frac{I}{N}, \quad r = \frac{R}{N}, \quad c = \frac{C}{N}, \quad s + i + r = 1.$$

For simplifications, we assume that at inception s = 1, r = c = 0 and compute the first order small variations $\Delta_i, \Delta_r, \Delta_c$. $\Delta_i(0)$ is the initial number of infected individuals.

(5) $\quad \frac{d\Delta_i}{dt} = (\beta - \gamma)\Delta_i, \quad \frac{d\Delta_r}{dt} = \gamma\Delta_i, \quad \frac{d\Delta_c}{dt} = \frac{\Delta_r}{\Delta_r + \Delta_i}\gamma\Delta_i,$

(6) $\Delta_i(t) = \Delta_i(0)e^{(\beta-\gamma)t}, \quad \Delta_r(t) = \frac{\gamma}{\beta-\gamma}\Delta_i(0)e^{(\beta-\gamma)t}, \quad \Delta_c(t) = \frac{\gamma^2}{\beta(\beta-\gamma)}\Delta_i(0)e^{(\beta-\gamma)t},$

(7) $\quad CCF = \frac{\Delta_r}{\Delta_r + \Delta_i} = \frac{\gamma}{\beta} = \frac{\Delta_c}{\Delta_r} = KSF.$

**SEIR model**. Using the same notations and assumptions for the SEIR model, one gets

(8) $\quad \frac{d\Delta_e}{dt} = \beta\Delta_i - \delta\Delta_e, \quad \frac{d\Delta_i}{dt} = \delta\Delta_e - \gamma\Delta_i, \quad \frac{d\Delta_r}{dt} = \gamma\Delta_i, \quad \frac{d\Delta_c}{dt} = \frac{\Delta_r}{\Delta_r + \Delta_i}\gamma\Delta_i,$

(9) $\Delta_i(t) = \Delta_i(0)e^{\lambda t}, \Delta_e(t) = \frac{\lambda + \gamma}{\delta}\Delta_i(0)e^{\lambda t}, \Delta_r(t) = \frac{\gamma}{\lambda}\Delta_i(0)e^{\lambda t}, \Delta_c(t) \frac{\gamma^2}{\lambda(\lambda+\gamma)}\Delta_i(0)e^{\lambda t},$

(10) $\quad CCF = KSF = \frac{\gamma}{\lambda + \gamma}, \quad \lambda = \frac{\sqrt{(\delta+\gamma)^2 + 4\delta(\beta-\delta)} - (\delta+\gamma)}{2}.$

**Simulations.** We performed discrete stochastic simulations of both SIR and SEIR models for different infectivity distributions, where each event is explicitly modeled. The models studied either had an equal probability of getting infected for each susceptible, or a variable distribution with a power-law distribution. We present here results with a slope of -2 in Fig. 2, but other slopes had similar results.

The simulations were performed as asynchronic stochastic realizations of the interactions described in Fig. 1. We compute the normalized probabilities of each type of event (infection or detection) in the appropriate model. At each step, we choose an event following these probabilities. For an infection event, a Susceptible is chosen based on its (pre-defined) infectivity. The probability of such an event is the product of the total number of Infected, the total infection probability of Susceptible individuals, and the infection rate $\beta$. For a detection event, an individual is randomly chosen with a probability proportional to the product of the total number of Infected and the detection rate $\gamma$.

Once the total number of diagnosed infected reaches one percent of the total population, we stop the simulation. The ratios in Fig. 2a, b, and d are taken along the simulations. The results in Fig. 2c, e, and f are at the last time point of the simulations. Simulations where the total number of Infected collapsed before reaching one percent of the total population were not incorporated in the results. Table 1 presents the average over 5 simulations for each set of parameters (corresponding to each point in Fig. 2c, e, and f) along with the analytical values.

| Model | β | δ | γ | 1 / KSF | 1 / CCF | Analytical |
|---|---|---|---|---|---|---|
| SIR | 1 | - | 0.3 | 3.36 | 3.34 | 3.33 |
|  |  |  | 0.5 | 2.15 | 2.04 | 2.00 |
|  |  |  | 0.7 | 1.45 | 1.41 | 1.43 |
| SIR | 2 | - | 0.3 | 6.71 | 6.53 | 6.67 |
|  |  |  | 0.5 | 3.94 | 3.97 | 4.00 |
|  |  |  | 0.7 | 2.81 | 2.90 | 2.86 |
| SIR | scale free | - | 0.3 | 2.80 | 2.76 | - |
|  |  |  | 0.5 | 1.77 | 1.77 | - |
|  |  |  | 0.7 | 1.38 | 1.34 | - |
| SEIR | 1 | 5 | 0.3 | 2.95 | 2.97 | 2.98 |
|  |  |  | 0.5 | 1.87 | 1.85 | 1.84 |
|  |  |  | 0.7 | 1.39 | 1.37 | 1.36 |
| SEIR | 2 | 5 | 0.3 | 5.34 | 5.29 | 5.30 |
|  |  |  | 0.5 | 3.26 | 3.27 | 3.26 |
|  |  |  | 0.7 | 2.40 | 2.37 | 2.39 |
| SEIR | 1 | 10 | 0.3 | 3.03 | 3.02 | 3.13 |
|  |  |  | 0.5 | 1.96 | 1.91 | 0.96 |
|  |  |  | 0.7 | 1.37 | 1.39 | 1.39 |
| SEIR | scale free | 5 | 0.3 | 2.53 | 2.57 | - |
|  |  |  | 0.5 | 1.63 | 1.64 | - |
|  |  |  | 0.7 | 1.26 | 1.26 | - |

**Table 1.** Average values obtained for *1 / CCF* and *1 / KSF* over 5 simulations for each set of parameters in SIR and SEIR models.

To compute the confidence intervals in Fig. 3, we ran simulations of the SIR models described above, and computed the ratio *KSF / CCF* as a function of the number of diagnosed infected. The mean of this ratio is 1 as expected and the standard deviation decreases below 10% after a minimal start-up (data not shown). We use this value as an upper bound of the standard deviation to compute the confidence intervals.

To improve the efficacy of the simulations, we used a tree formalism[38]. Each event type (infection or detection) is represented as a tree to allow a rapid selection of the individual involved in the next event. Each leaf corresponds to an individual. The value of each internal node in the tree is the sum of the values in its direct descendants in the tree. The tree root is the total probability of the event. This configuration enables us to access each individual in a logarithmic time. In case of an infection event, a Susceptible becomes Infected, the chosen Susceptible is determined by traversing the Susceptible tree. The tree is then updated along the entire path. We also randomly choose an Infected as the contaminator and record its leaf number in a repertoire. For a detection event, an individual is randomly chosen in the Infected tree. We then check if the contaminator has already been detected by observing if the leaf of the contaminator was already detected. The number of Controlled is then increased by 1.

## Data availability

The code used and the datasets generated and/or analyzed during the current study are available from the corresponding author on reasonable request and can also be found at https://github.com/louzounlab/undiagnosed_fraction_estimation

## References

1  Squazzoni, F., et al. Computational models that matter during a global pandemic outbreak: A call to action. *Journal of Artificial Societies and Social Simulation* **23** (2020).

2  Prem, K., Liu, Y., Russell, T.W. et al. The effect of control strategies to reduce social mixing on outcomes of the COVID-19 epidemic in Wuhan, China: a modelling study. *Lancet Public Health* **5,** e261-70 (2020).

3  Tian, H., Liu, Y., Li. Y. et al. An investigation of transmission control measures during the first 50 days of the COVID-19 epidemic in China. *Science* **368**, 638—42 (2020).

4  Smith, R. D. Responding to global infectious disease outbreaks: lessons from SARS on the role of risk perception, communication, and management. *Social science & medicine* **63,** 3113-3123 (2006).

5  Punyacharoensin, N., et al. Modelling the HIV epidemic among MSM in the United Kingdom: quantifying the contributions to HIV transmission to better inform prevention initiatives. *Aids* **29,** 339—349 (2015).

6  Glass, K., Becker, N., & Clements, M. Predicting case numbers during infectious disease outbreaks when some cases are undiagnosed. *Statistics in Medicine* **26,** 171—183 (2007).

7  Christaki, E. New technologies in predicting, preventing, and controlling emerging infectious diseases. *Virulence* **6,** 558—565 (2015).

8  Hutchinson, S. J., et al. Method used to identify previously undiagnosed infections in the HIV outbreak at Glenochil prison. *Epidemiology & Infection* **123,** 271—275 (1999).

9  Britton, T., & Scalia Tomba, G. Estimation in emerging epidemics: Biases and remedies. *Journal of the Royal Society Interface* **16,** 20180670 (2019).

10  Yuen, K.S., Ye, Z.W., Fung, S.Y., Chan, C.P., and Jin, D.Y. SARS-CoV-2 and COVID-19: The most important research questions. *Cell & bioscience* **10**, 1—5 (2020).

11 Grant, A. Dynamics of COVID-19 epidemics: SEIR models underestimate peak infection rates and overestimate epidemic duration. Preprint at https://www.medrxiv.org/content/medrxiv/early/2020/04/06/2020.04.02.20050674.full.pdf (2020).

12 Richterich, P. Severe underestimation of COVID-19 case numbers: effect of epidemic growth rate and test restrictions. Preprint at https://www.medrxiv.org/content/medrxiv/early/2020/04/17/2020.04.13.20064220.full.pdf (2020).

13 Chowell, G., et al. Model parameters and outbreak control for SARS. *Emerging infectious diseases* **10,** 1258 (2004).

14 Viboud, C., Simonsen, L., & Chowell, G. A generalized-growth model to characterize the early ascending phase of infectious disease outbreaks. *Epidemics* **15,** 27—37 (2016).

15 Finkenstädt, B. F., Bjørnstad, O. N., & Grenfell, B. T. A stochastic model for extinction and recurrence of epidemics: estimation and inference for measles outbreaks. *Biostatistics* **3,** 493—510 (2002).

16 Theagarajan, L.N. Group testing for COVID-19: how to stop worrying and test more. Preprint at https://arxiv.org/pdf/2004.06306.pdf (2020).

17 Flaxman, S., Mishra, S., Gandy, A., et al. Estimating the number of infections and the impact of non-pharmaceutical interventions on COVID-19 in 11 European countries. Preprint at https://arxiv.org/pdf/2004.11342.pdf (2020).

18 Zhigljavsky, A., Whitaker, R., Fesenko, I. et al. Generic probabilistic modelling and non-homogeneity issues for the UK epidemic of COVID-19. Preprint at https://arxiv.org/pdf/2004.01991.pdf (2020).

19 Wearing, H.J., Rohani, P. & Keeling, M.J. Appropriate models for the management of infectious diseases. *PLoS medicine* **2**,174 (2005).

20 Giudici, M., Comunian, A. & Gaburro, R. Inversion of a SIR-based model: a critical analysis about the application to COVID-19 epidemic. Preprint at https://arxiv.org/pdf/2004.07738.pdf (2020).

21 Cooper, I., Mondal, A., Antonopoulos, C.G. A SIR model assumption for the spread of COVID-19 in different communities. *Chaos, Solitons & Fractals* **139,** 110057 (2020).

22 Yan, G., Lee, C.K., Lam, L.T. et al. Covert COVID-19 and false-positive dengue serology in Singapore. *Lancet Infectious Diseases* **20**, 536 (2020).

23 Pandey, G., Chaudhary, P., Gupta, R. & Pal, S. SEIR, and Regression Model based COVID-19 outbreak predictions in India. Preprint at https://arxiv.org/ftp/arxiv/papers/2004/2004.00958.pdf (2020).

24 Chinazzi, M., Davis, J.T., Ajelli, M. et al. The effect of travel restrictions on the spread of the 2019 novel coronavirus (COVID-19) outbreak. *Science* **368**, 395—400 (2020).

25 Singh, R. & Adhikari, R. Age-structured impact of social distancing on the COVID-19 epidemic in India. Preprint at https://arxiv.org/pdf/2003.12055.pdf (2020).

26 Bai, Y., Yao, L., Wei, T., Tian, F., Jin, D.Y., Chen, L. & Wang, M. Presumed asymptomatic carrier transmission of COVID-19. *Jama* 323, 1406—7 (2020).

27 Berger, D.W., Herkenhoff, K.F. & Mongey, S. An seir infectious disease model with testing and conditional quarantine. *National Bureau of Economic Research* (2020).

28    Hsieh, Y. H. et al. Impact of quarantine on the 2003 SARS outbreak: a retrospective modeling study. *Journal of Theoretical Biology*, **244,** 729—736. (2007).

29    Manski, C.F., Molinari, F. Estimating the COVID-19 infection rate: Anatomy of an inference problem. *Journal of Econometrics* **220,** 181-192 (2020).

30    Villalobos, C. SARS-CoV-2 infections in the World: An estimation of the infected population and a measure of how higher detection rates save lives. *Frontiers in public health*, **8,** 489 (2020).

31    Kemper, J.T. On the identification of superspreaders for infectious disease. *Mathematical Biosciences*, **48**, 111—27 (1980).

32    Wong, G. et al.  MERS, SARS, and Ebola: the role of super-spreaders in infectious disease. Cell host & microbe. Cell host & microbe, **18,** 398—401 (2015).

33    Mueller, M., Derlet, P.M., Mudry, C. & Aeppli, G. Using random testing to manage a safe exit from the COVID-19 lockdown. Preprint at https://arxiv.org/pdf/2004.04614.pdf (2020).

34    Pollán, M., Pérez-Gómez, B., Pastor-Barriuso, R. et al. Prevalence of SARS-CoV-2 in Spain (ENE-COVID): a nationwide, population-based seroepidemiological study. *The Lancet*, **396,** 535—44 (2020).

35    Nishiura, H., et al. Estimation of the asymptomatic ratio of novel coronavirus infections (COVID-19). *International journal of infectious diseases* **94,** 154 (2020).

36    Kretzschmar M.E. et al. Impact of delays on effectiveness of contact tracing strategies for COVID-19: a modelling study. *The Lancet Public Health* **5,** e452-9 (2020).

37    Park, Y.J. et al. Contact tracing during coronavirus disease outbreak, South Korea, 2020. *Emerging infectious diseases*, **262,** 465—8 (2020).

38    Melka, A., Dori, N., and Louzoun Y. Invasion rate versus diversity in population dynamics with catastrophes. *Physical Review Letters*, **124,** 158301 (2020).

## Authors contributions

All authors contributed equally.

## Competing interests

The authors declare no competing interests.

## Additional information

**Funding:** DoD Grant N629091912097